\documentclass{PoS}
\pdfoutput=1

\newcommand{\matrixel}[3]{\left< #1 \vphantom{#2#3} \right| #2 \left| #3 \vphantom{#1#2} \right>}
\newcommand{\avg}[1]{\left< #1 \right>}

\title{Domain Wall Charm Physics with Physical Pion Masses: Decay constants, bag and $\xi$ parameters}

\ShortTitle{Domain Wall Charm Physics with Physical Pion Masses}

\author{Peter Boyle, Luigi Del Debbio, Ava Khamseh\textsuperscript{*}\\
        School of Physics and Astronomy, University of Edinburgh\\
        EH9 3JZ, Edinburgh, United Kingdom\\
        E-mail: \email{a.khamseh@sms.ed.ac.uk}}
\author{Andreas J\"uttner, Francesco Sanfilippo, Justus Tobias Tsang\textsuperscript{*}\\
        School of Physics and Astronomy, University of Southampton\\
        SO17 1BJ Southampton, United Kingdom\\
        E-mail: \email{j.t.tsang@soton.ac.uk}
        \phantom{\speaker{A. Khamseh, J. T. Tsang}}}
\author{RBC and UKQCD Collaborations}

\abstract{We provide an overview of RBC/UKQCD's charm project on 2+1 flavour physical pion mass ensembles using M\"obius Domain Wall Fermions for the light as well as for the charm quark. We discuss the analysis strategy in detail and present results at the different stages of the analysis for $D$ and $D_s$ decay constants as well as the bag and $\xi$ parameters. We also discuss future approaches to extend the reach in the heavy quark mass.}

\FullConference{The 33rd International Symposium on Lattice Field Theory\\
		14 -18 July 2015\\
		Kobe International Conference Center, Kobe, Japan}

\begin{document}

\section{Introduction}
In this work we present preliminary results of RBC/UKQCD's charmed meson project with $N_f=2+1$ dynamical domain wall fermions. Our current charm physics program aims to give a fully controlled prediction of $D$ and $D_s$ decay constants as well as the bag and $\xi$ parameters. We use the Iwasaki gauge action~\cite{Iwasakiaction} and the domain wall fermion action~\cite{Brower:2004xi,Shamir:1993zy,Furman:1994ky} with M\"obius-kernel \cite{Blum:2014tka} and explore the charmed mesons using three different lattice spacings with a range of pion and charm masses. This is done at two lattice spacings with near-physical pion masses and at third lattice spacing at a higher pion mass. The simulations are also carried out on four further ensembles with heavier pion masses which are needed for the extrapolation to physical pion masses and allowing for a continuum extrapolation with three lattice spacings. The goal is to extrapolate observables obtained from the simulated data to $D$ and $B$ mesons (using the ETMC ratio method~\cite{Blossier:2009hg}) after performing the physical pion mass and continuum extrapolations. This proceeding is a preliminary status report demonstrating the quality of our data and outlining the analysis strategy employed.

The first part of the analysis focusses on the determination of the decay constants $f_D$ and $f_{D_s}$ defined by
\begin{equation}
\matrixel{0}{A_{cq}^\mu}{D_q(p)} = f_{D_q}p^\mu_{D_q},
\end{equation}
where $q=d,s$ and the axial current is given by $A^\mu_{cq} = \bar{c}\gamma_\mu \gamma_5 q$. Precise knowledge of $f_D$ and $f_{D_s}$ together with experimental input of the measured branching ratios $\mathcal{B}(D_{(s)} \to l \nu_l)$ and total width $\Gamma$ allow for the determination of $|V_{cd}|$ and $|V_{cs}|$ respectively. Combined with calculations of other CKM matrix elements~\cite{Aoki:2013ldr}, this in turn provides a test of the unitarity of the CKM matrix and therefore a test of the Standard Model.

The second part of the analysis focuses on computing non-perturbative short distance contribution to neutral meson mixing. Experimantally, $B$ meson mixing is observed in terms of oscillation frequencies $\Delta m_q$, which are conventionally parameterised by 
\begin{equation}\label{eqn:deltampheno}
\Delta m_q=\frac{G_F^2m_W^2}{16\pi^2m_{B_q}}|V_{tq}^*V_{tb}^*|S_0(x_t)\eta_B\langle\bar{B}_q^0|\left[\bar{b}\gamma^\mu(1-\gamma_5)q\right]\left[\bar{b}\gamma^\mu(1-\gamma_5)q\right]|B_q^0\rangle.
\end{equation}
Here $q$ stands for either a $d$ or an $s$ quark. The Inami-Lim function $S_0(x_t)$ and $\eta_B$ can be computed using perturbation theory \cite{Albertus:2010nm}.\\ 

\noindent The non-perturbative contribution to neutral meson mixing due to weak interactions is given by the matrix element in Equation \ref{eqn:deltampheno} which we compute on the lattice in terms of the bag parameter 

\begin{equation}
\label{eqn:bagparam}
B_P=\frac{\langle P^0|O_{VV+AA}|\bar{P}^0\rangle}{\frac{8}{3}f_P^2m_P^2},
\end{equation}
with the four-quark operator 
\begin{equation}
O_{VV+AA}=(\bar{h}\gamma_\mu q)(\bar{h}\gamma_\mu q)+(\bar{h}\gamma_5\gamma_\mu q)(\bar{h}\gamma_5\gamma_\mu q).
\end{equation}
Our strategy in this project is to compute the bag parameter in the charm mass region and then extrapolate to the $b$ quark mass using the ETMC ratio method. In addition, we will also apply the ratio method in order to determine the parameter $\xi$, given by the ratio of $B_s$ meson mixing over $B_d$ meson mixing

\begin{equation}
\xi=\frac{f_{hs}\sqrt{B_{hs}}}{f_{hl}\sqrt{B_{hl}}}.
\end{equation}

\noindent When combined with the experimental measurements, this quantity allows to extract the ratio of CKM matrix elements $|V_{td}/V_{ts}|$ which enters as an important constraint in fits of the unitarity \linebreak triangle.  \\

\noindent For chiral fermions, there is no operator mixing in the Standard Model and only one 4-quark \linebreak operator, i.e. $O_{VV+AA}$, contributes. Also, the renormalization factor for the bag parameter cancels between the numerator and the denominator. As a result no renormalization is required for $\xi$ when the actions used for both heavy and light quarks have chiral symmetry.  


\section{Ensembles and Measurement Parameters} \label{sec:ensembles}
\begin{table}
  \begin{center}
    \begin{tabular}{c c c c c c c c}
      \hline
      Name & $L/a$ & $T/a$ &  $a^{-1}[\mathrm{GeV}]$ & $m_\pi[\mathrm{MeV}$] & hits/conf & confs & total\\\hline
      C0  & 48     & 96     & 1.73   & 139  & 48 & 88  & 4224 \\
      C1  & 24     & 64     & 1.78   & 340  & 32 & 100 & 3200 \\
      C2  & 24     & 64     & 1.78   & 430  & 32 & 101 & 3232 \\\hline
      M0  & 64     & 128    & 2.36   & 139  & 32 & 80  & 2560 \\
      M1  & 32     & 64     & 2.38   & 300  & 32 & 72  & 2304 \\
      M2  & 32     & 64     & 2.38   & 360  & 16 & 76  & 1216 \\\hline
      F1  & 48     & 96     & $\sim$2.77$\phantom{\sim}$   & 230  & 48 & 70  & 3360 \\ \hline\hline
      A1  & 16     & 32     & 1.78   & 430  &  - & -   & -    \\
      \hline 
    \end{tabular}
  \end{center}
  \caption{This table summarises the main parameters of our $N_f=2+1$ ensembles. C stands for coarse, M for medium and F for fine. A1 corresponds to an auxiliary ensemble we used to test envisaged future approaches. The lattice spacing on F1 is still preliminary.}
  \label{tab:ensembles}
  \end{table}

The simulations were carried out on three different lattice spacings ({\bf C}oarse, {\bf M}edium and {\bf F}ine) with a number of different pion masses. The ensemble details can be found in Table \ref{tab:ensembles}. In particular this includes RBC-UKQCD's physical pion mass M\"obius domain wall ensembles ({\bf C0} and {\bf M0})~\cite{Blum:2014tka}. To control the continuum limit we additionally generated a finer ensemble {\bf F1} with a pion mass of $m_\pi\approx 230\, \mathrm{MeV}$. To be able to carry out the extrapolation to physical pion masses on this ensemble we also simulated on RBC-UKQCD's Shamir domain wall fermion ensembles~\cite{Allton:2008pn,Allton:2007hx} with (nearly) the same lattice spacings as {\bf C0} and {\bf M0}. The {\bf A}uxiliary ensemble {\bf A1}~\cite{Allton:2007hx} was only used to test future approaches as will be discussed in Section \ref{sec:gaugesm}.

\begin{table}
  \begin{center}
    \begin{tabular}{c c c c c c c c c}
      \hline
      Name & DWF & $M_5$  & $L_s$ & $am_l^\mathrm{uni}$ & $am_s^{\mathrm{uni}}$ & $am_s^{\mathrm{sim}}$ & $am_s^{\mathrm{phys}}$ & $\Delta m_s/m_s^{\mathrm{phys}}$\\\hline
      C0  & MDWF    & 1.8    &  24   & 0.00078  & 0.0362   & 0.0362           & 0.0358  & 1.1\% \\
      C1  & SDWF    & 1.8    &  16   & 0.005    & 0.04     & 0.03224, 0.04    & 0.03224 & 0.0\%\\
      C2  & SDWF    & 1.8    &  16   & 0.01     & 0.04     & 0.03224          & 0.03224 & 0.0\% \\\hline
      M0  & MDWF    & 1.8    &  12   & 0.000678 & 0.02661  & 0.02661          & 0.02539 & 4.8\% \\
      M1  & SDWF    & 1.8    &  16   & 0.004    & 0.03     & 0.02477, 0.03    & 0.02477 & 0.0\% \\
      M2  & SDWF    & 1.8    &  16   & 0.006    & 0.03     & 0.02477          & 0.02477 & 0.0\% \\\hline
      F1  & MDWF    & 1.8    &  12   & 0.002144 & 0.02144  & 0.02144          & -       & - \\
      \hline 
    \end{tabular}
  \end{center}
  \caption{Domain wall parameters for the light and strange quarks of all ensembles.  All quoted values for $am_l$ and $am_s$ are bare quark masses in lattice units. The column DWF corresponds to the chosen domain wall fermion formulation where 'MDWF' corresponds to M\"obius domain wall fermions, 'SDWF' to Shamir domain wall fermions. All light quarks were simulated at their unitary value $am^\mathrm{uni}$. Valence strange quarks were simulated at $am_s^\mathrm{sim}$. Note that the value of the physical strange quark mass $am_s^\mathrm{phys}$ slightly disagrees with the unitary strange quark mass $am_s^\mathrm{uni}$.}
  \label{tab:DWFls}
\end{table}

For the light and strange sector we simulate the unitary light quark masses and only slightly adjust the strange quark mass (compare Table \ref{tab:DWFls}) to simulate directly at its physical value as stated in ~\cite{Blum:2014tka}. Changing the action from Shamir domain wall fermions ({\bf C1}, {\bf C2}, {\bf M1}, {\bf M2}), to M\"obius domain wall fermions ({\bf C0}, {\bf M0}, {\bf F1}) allows for a significant reduction of the extent of the fifth dimension $L_s$, thus reducing the computational cost. However, the two actions are chosen to lie on approximately the same scaling trajectory, therefore a combined continuum limit can be taken. The choice of all the light quark parameters entering the simulation is given in Table \ref{tab:DWFls}.

The simulation of charm quarks using a domain wall action is challenging.
To show that domain wall fermions are a suitable discretisation for charm phenomenology, we carried out a quenched pilot study~\cite{Cho:2015ffa,Tsang:2014pfa}.
In this pilot study we mapped out the parameter space of the domain wall action to optimise it for the simulation of charm quarks. 
We found that a domain wall height $M_5=1.6$ and an extend of the 5th dimension of $L_s=12$ was the optimal choice to simulate heavy quarks with a M\"obius domain wall formalism. This works reliably for $am_h \lesssim 0.4$, so in the following we restricted ourselves to input quark masses in bare lattice units of $am_h \leq 0.4$ as can be seen in Table \ref{tab:DWFh}~\cite{Tsang:2014pfa}. The only exception to this is the input quark mass of $am_h = 0.45$ on {\bf C0} which is indicated in red in Table \ref{tab:DWFh}. With this we tested the reach in the heavy quark mass of our formulation on dynamical 2+1 flavour ensembles. This data point does not enter any of the subsequent analyses and is only shown as an open symbol. Our assumption that the qualitative features of the quenched pilot study remain the same in the dynamical case has so far been confirmed~\cite{Boyle:2015rka}.

\begin{table}
  \begin{center}
    \begin{tabular}{c c c c c}
      \hline
      Name & $L_s$ & $am_h^{\mathrm{bare}}$         \\\hline
      C0   & 1.6    &  12   & 0.3,  0.35, 0.4, {\color{red} 0.45}  \\
      C1   & 1.6    &  12   & 0.3,  0.35, 0.4             \\
      C2   & 1.6    &  12   & 0.3,  0.35, 0.4             \\\hline
      M0   & 1.6    &  12   & 0.22, 0.28, 0.34, 0.4       \\
      M1   & 1.6    &  12   & 0.22, 0.28, 0.34, 0.4       \\
      M2   & 1.6    &  12   & 0.22, 0.28, 0.34, 0.4       \\\hline
      F1   & 1.6    &  12   & 0.18, 0.23, 0.28, 0.33, 0.4 \\
      \hline 
    \end{tabular}
  \end{center}
  \caption{M\"obius domain wall parameters for the heavy quarks of all ensembles. All quoted values for $am_h$ are bare quark masses in lattice units. As described in the text, the value indicated in red was only used to verify our assumptions about the applicability of the quenched pilot study to the dynamical case.}
  \label{tab:DWFh}
\end{table}

There are several factors allowing us to achieve the presented precision: we use $\mathbb{Z}(2) \times \mathbb{Z}(2)$ stochastic sources \cite{Boyle:2008rh} on a large number of time planes (compare hits/conf column in Table \ref{tab:ensembles}), giving rise to a stochastic estimate of the $L^3\times N_\mathrm{hits}$ translational volume average of the correlation functions.
 The operator inversions are then performed using the HDCG algorithm \cite{Boyle:2014rwa} for light and strange propagators, reducing the numerical cost and hence making this computation feasible. For the heavy quark propagators a CG inverter is used.

\section{Decay Constant Analysis}
To make a prediction for the decay constants $f_D$ and $f_{D_s}$ with fully controlled systematics, a number of inter- and extrapolations need to be performed on the data. First, masses and matrix elements (and hence decay constants) are obtained by fitting the correlation functions to obtain $\mathcal{O}(a,m_l,m_h,m_s)$ where $\mathcal{O}$ can be any observable containing at least one heavy quark, e.g. the pseudoscalar mass $m_P$, its decay constant $f_P$ or the quantity $\Phi_P = f_P \sqrt{m_P}$ (or ratios of these), where $P=D$, $D_s$ or $\eta_c$. On {\bf C0} and {\bf M0} we correct quantities with a valence strange quark for the mistuning in $am_s$ that is present on those ensembles (c.f. the last column in Table \ref{tab:DWFls}) giving $\mathcal{O}(a,m_l,m_h,m_s^\mathrm{phys})$. Subsequently, the heavy quark dependence is fixed by extrapolating the data to common reference masses for some choice of meson including at least one valence heavy quark and no valence $u/d$ quarks (i.e. either $m_{D_s}$ or $m_{\eta_c}$). This is done individually for each ensemble and one obtains $\mathcal{O}(a,m_l,m_{D_s/\eta_c}^\mathrm{ref},m_s^\mathrm{phys})$. This data then undergoes a very small extrapolation to physical pion masses with some additional assumptions guiding the extrapolation for {\bf F1}, yielding $\mathcal{O}(a,m_\pi^\mathrm{phys},m_{D_s/\eta_c}^\mathrm{ref},m_s^\mathrm{phys})$. Cut-off effects are removed by taking the continuum limit giving $\mathcal{O}(a=0,m_\pi^\mathrm{phys},m_{D_s/\eta_c}^\mathrm{ref},m_s^\mathrm{phys})$. Finally, we interpolate the obtained data for the different $m_{D_s/\eta_c}^\mathrm{ref}$ to obtain the physical value $\mathcal{O}(a=0,m_\pi^\mathrm{phys},m_{D_s/\eta_c}^\mathrm{phys},m_s^\mathrm{phys})$. For the determination of the errors in this entire analysis the bootstrap resampling method with 2000 bootstrap samples was used.

\subsection{Correlation Function Fits and Mass Interpolations}
The fits leading to the extraction of masses and decay constants are simultaneous multi-channel fits to the two-point correlation functions $\avg{AA}$, $\avg{AP}$, $\avg{PA}$ and $\avg{PP}$ where $P$ is the pseudoscalar operator and $A$ is the operator for the temporal component of the axial current. Correlated fits were carried out by thinning the correlation matrix (i.e. only including every $n$th time slice in the fit), whilst monitoring the condition number of the correlation matrix to ensure numerical convergence. To increase the statistical precision we fitted the ground state as well as the first excited state, allowing for earlier time slices (with smaller statistical errors) to contribute to the fit. During all these fits the $\chi^2/\mathrm{d.o.f}$ and the corresponding $p$-values were monitored, ensuring that the $p$-values are always above $5\%$.

\begin{figure}
  \center
  \includegraphics[width=\textwidth]{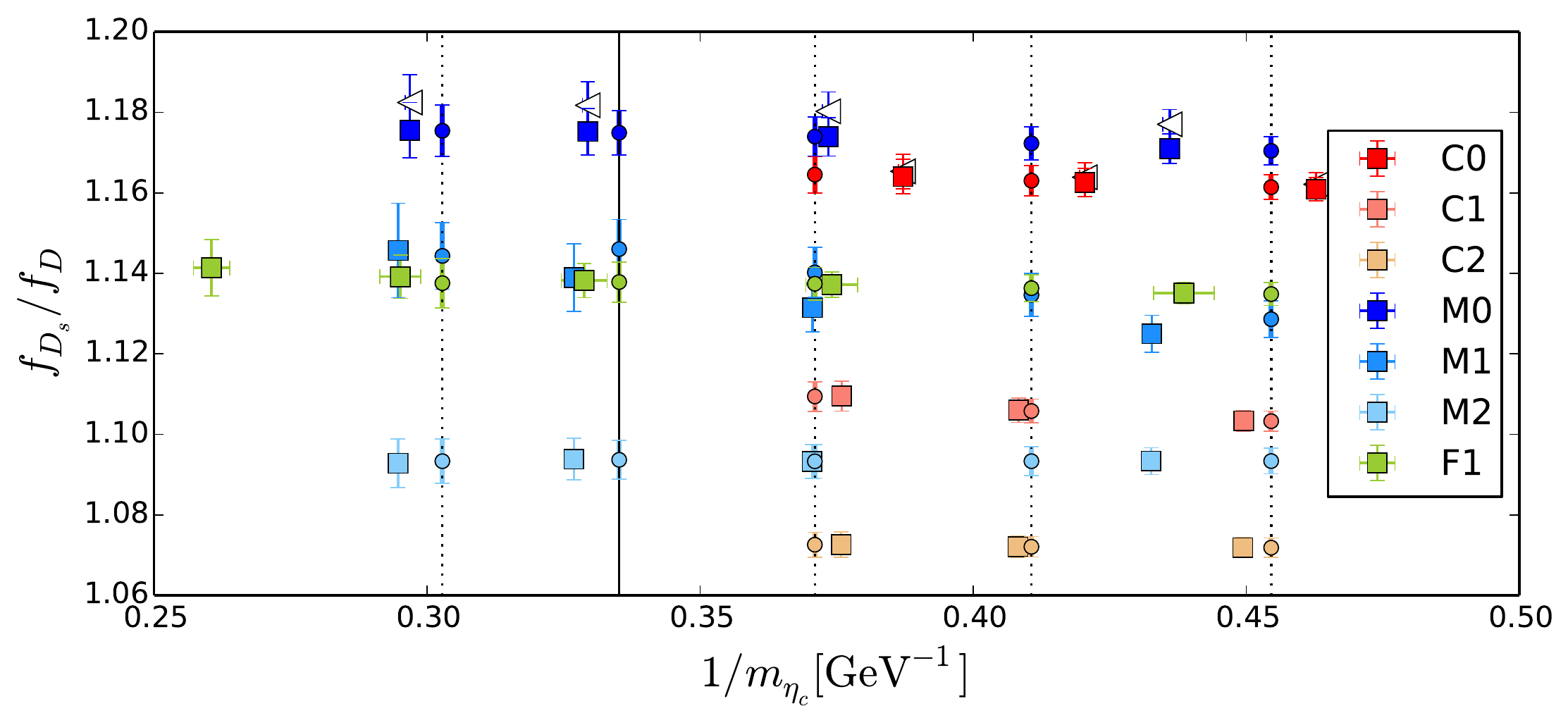}
  \caption{Collected data as obtained from correlator fits. Because the calculation of the renormalization constants $Z_A$ is still in progress, we only quote that ratio of decay constants $f_{D_s}/f_D$ here. The dotted vertical lines correspond to the reference masses $m_{\eta_c}^{\mathrm{ref},i}$, the solid vertical line corresponds to the physical $\eta_c$ mass. The open triangles correspond to the simulated data with mistuned valence strange quark masses, the squares to the data after correcting for the mistuned valence strange quark mass. The filled circles on the vertical lines are the results of the interpolation to the respective reference masses.}
  \label{fig:datacollection}
\end{figure}

As stated above, the ensembles {\bf C0} and {\bf M0} have slightly mistuned strange quark masses~\cite{Blum:2014tka}. We correct for the mistuned valence strange quark mass, the effect of the mistuning of the strange quark mass in the sea is assumed to be small. Once the analysis is finalised we will estimate the effect of the sea quark mistuning and include it as a systematic error. To this end we simulated not just the physical valence strange quark mass but also the unitary one on {\bf C1} and {\bf M1}, to obtain some information of the needed corrections on {\bf C0} and {\bf M0}, respectively. This was done for the same heavy quark masses $am_h^i$ as described in Table \ref{tab:DWFh}. From these partially quenched data points we deduce dimensionless parameters $\alpha_\mathcal{O}(am_h^i)$ for the different observables $\mathcal{O}$ using
\begin{equation}
\mathcal{O}^\mathrm{phys} = \mathcal{O}^\mathrm{uni} \left(1 + \alpha_\mathcal{O} \frac{\Delta m_s}{m_s^\mathrm{phys}} \right),
\label{eq:definition_alpha}
\end{equation}
where $\Delta m_s \equiv m_s^\mathrm{uni} - m_s^\mathrm{phys}$.
The definition of these $\alpha_\mathcal{O}(am_h^i)$ ensures that they are independent of renormalization constants and the lattice spacing. This was done for all quantities that contain a strange valence quark, i.e. $\mathcal{O} = m_{D_s}$, $f_{D_s}$ and $f_{D_s}\sqrt{m_{D_s}}$ and for each choice of the simulated heavy quark mass. The maximum absolute value of $\alpha_\mathcal{O}(am_h^i)$ (varying the heavy quark masses as well as the observables) is $|\alpha| \leq 0.15$. So the size of the maximal correction to the data is less than $1\%$. 
Figure \ref{fig:datacollection} presents the results of the correlator fits for $f_{D_s}/f_D$. For the case of {\bf C0} and {\bf M0} the open triangles indicate the data before the correction has been applied. In the case of {\bf C0} this correction is very small, so the red squares effectively lie on top of th eopen triangles. For all ensembles, the filled squares correspond to the values for $m_s = m_s^\mathrm{phys}$. The effect of the strange quark correction is illusatrated by the shift from the open triangles to the filled squares on these two ensembles.

Eventually we want to carry out an extrapolation to physical pion masses and take the continuum limit of the results. We need to ensure that the data lie on a line of constant physics and therefore that all quantities are considered at the same heavy quark mass. This is done by interpolating the data to common reference meson masses $m^\mathrm{ref}_P$ where the meson $P$ includes a heavy quark, i.e. either $m_{\eta_c}$ or $m_{D_s}$. We choose to carry out the interpolation in $1/m_{\eta_c}$. This is motivated by the fact that such a meson does not include a valence light or strange quark, thus it is not affected by any interpolation to physical light and strange valence quark masses. In the following $\eta_c$ corresponds to its connected part only.
The interpolation is simply done as a linear spline, but to keep control of the systematic error associated with this interpolation, we also carry out the interpolation as a quadratic spline and as a global second order polynomial. The maximum spread of the central values is then added in quadrature to the statistical error. We also interpolate the ratios of decay constants using the same method. 
We choose linearly spaced values for $1/m_{\eta_c}^{\mathrm{ref},i}$.
Where possible, we choose the spacing such that we have data on all ensembles in the region of the reference masses. Finally, we enforce that for one value we have $m_{\eta_c}^\mathrm{ref}|_\mathrm{conn.}=m_{\eta_c}^\mathrm{phys}= 2983.6(6)\mathrm{MeV}$~\cite{pdg-2015}. The interpolated values for the example of $\mathcal{O} = f_{D_s}/f_D$ are shown as filled circles in Figure \ref{fig:datacollection}.
The figure also reveals that we cannot simulate physical charm quark masses on the coarse ensembles obeying the condition $am_h\leq 0.4$~\cite{Tsang:2014pfa}.

\subsection{Chiral and Continuum Extrapolation}
After the previous interpolations and corrections we have obtained $\mathcal{O}(a,m_l,m_{D_s/\eta_c}^{\mathrm{ref},i},m_s^\mathrm{phys})$ for $\mathcal{O}=f_P$, $\Phi_P$ and their ratios, where $P=D,\, D_s$. In this step of the analysis we extrapolate to physical light quark masses by enforcing that $m_\pi=m_\pi^\mathrm{phys}$ and to vanishing lattice spacing, i.e. $a=0$.
So far the analysis can be carried out ensemble by ensemble avoiding the need of renormalization factors. In general, the chiral extrapolation and the continuum limit requires to renormalize the observables. Since we use a different discretisation for the light quarks than for the heavy quarks, we have to renormalize a mixed action current. We are in the process of determining this mixed action current renormalization non-perturbatively and this will be reported in future. For now we simply use the renormalization constants from the light-light current. This implies that the results presented here only serve to show the quality of our data and not as prediction for physical constants.

\begin{figure}
\center
\includegraphics[width=0.75\textwidth]{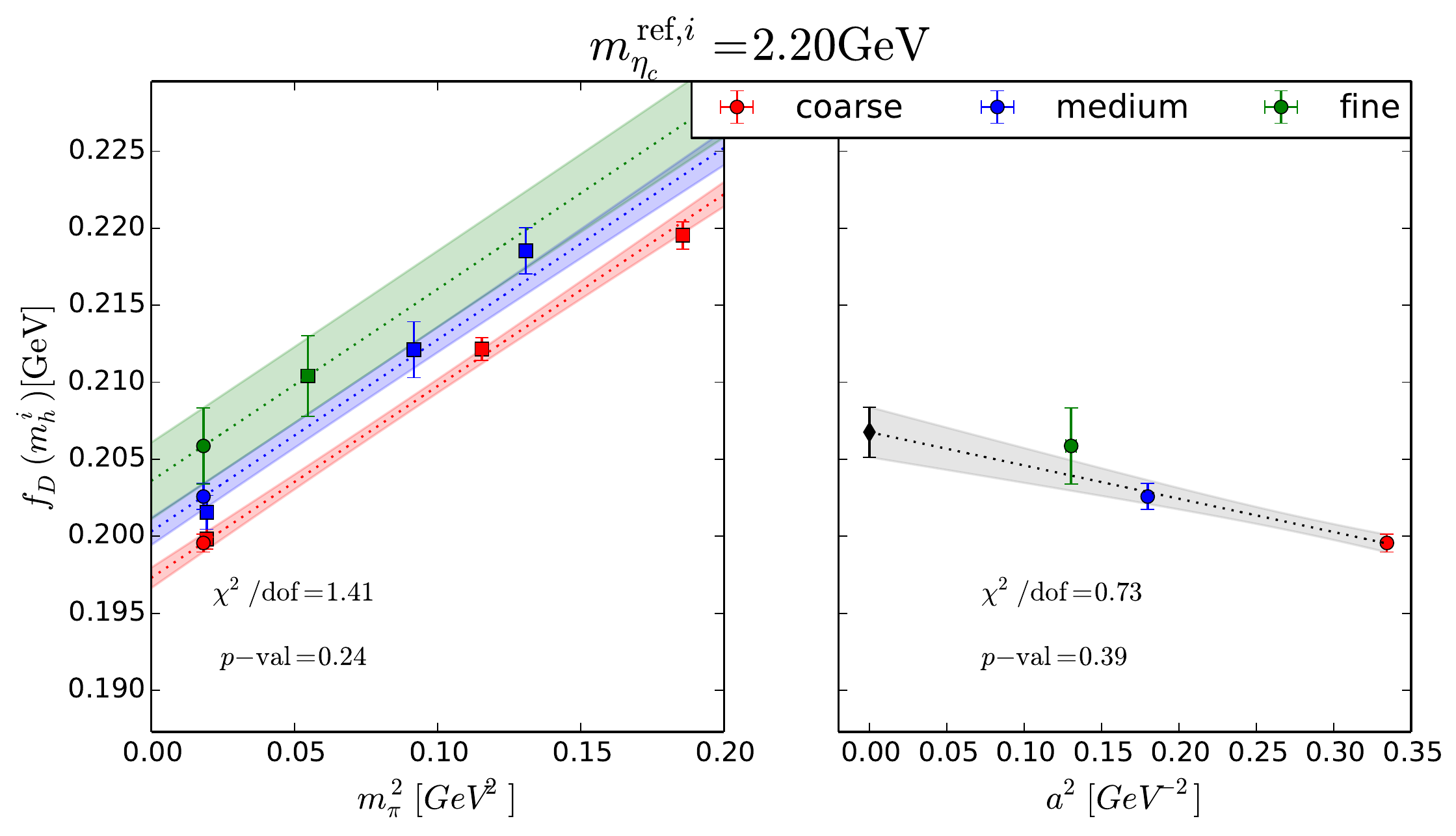}\\
\includegraphics[width=0.75\textwidth]{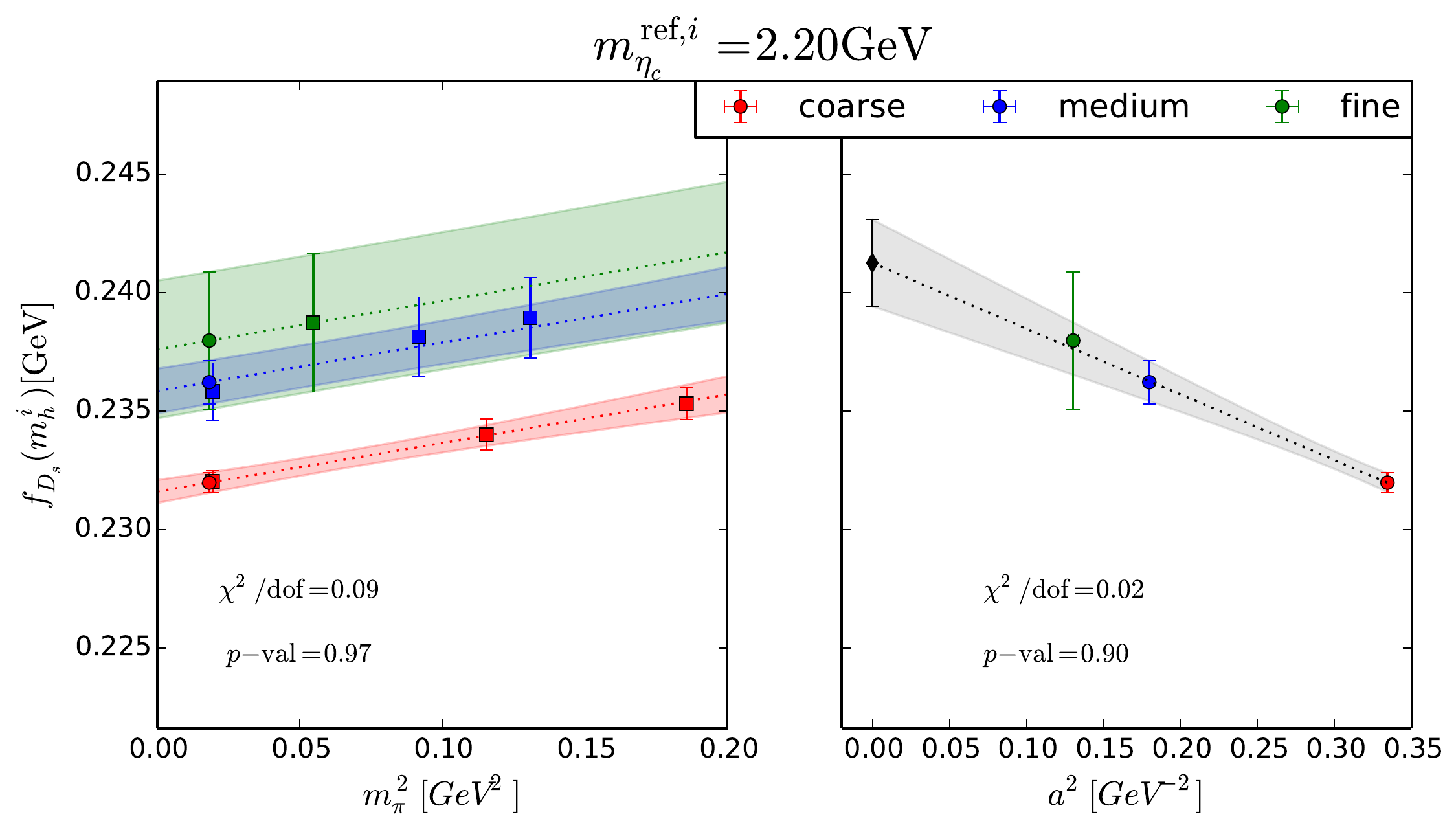}
\caption{Chiral and continuum extrapolation for $f_D$ (top) and $f_{D_s}$ (bottom) at a reference mass of $\eta_c=2.20\mathrm{GeV}$.
  In each plot the left panel is the extrapolation to physical pion masses whilst the right panel is the continuum limit from the obtained values at $m_\pi = m_\pi^\mathrm{phys}$.
  The large error bars of the fine ensemble arise, at this point, from the uncertainty of the exact lattice spacing.
  As mentioned before, this data is not yet correctly renormalized, so this figure serves as a presentation of our method and the quality of our data only.
  From the chiral extrapolation we find a cut-off independent slope for the coarse and the medium ensembles, which is then applied to the fine ensemble {\bf F1}. The exact details of this method are described in the text.}
\label{fig:chiralCL}
\end{figure}

For the ensembles {\bf C0} and {\bf M0} only a tiny extrapolation is required to reach physical pion masses. However, for the case of {\bf F1} the extrapolation extends further, so some assumption about how to carry out this extrapolation is needed. We parameterise the slope in $m_\pi^2$ in the following way:
\begin{equation}
  \mathcal{O}(a,m_\pi,m_{D_s/\eta_c}^{\mathrm{ref},i},m_s^\mathrm{phys}) =   \mathcal{O}(a,m_\pi=0,m_{D_s/\eta_c}^{\mathrm{ref},i},m_s^\mathrm{phys}) + C^\chi_\mathcal{O}(m_{D_s/\eta_c}^{\mathrm{ref},i})\, m_\pi^2.
  \label{eq:chiral}
\end{equation}
For each reference mass $m_{D_s/\eta_c}^{\mathrm{ref},i}$ the slope $C^\chi_\mathcal{O}$ of the chiral extrapolation is found by simultaneously fitting all coarse ({\bf C0}, {\bf C1}, {\bf C2}) and all medium ({\bf M0}, {\bf M1}, {\bf M2}) ensembles. This assumes that the slope is independent of cut-off effects and the fit quality can be tested by monitoring the $\chi^2/\mathrm{d.o.f}$ and the corresponding $p$-values, as stated in the plots of Figure \ref{fig:chiralCL}. Since the $p$-values are satisfactory (i.e. $p>0.05$), this slope is applied to {\bf F1} to obtain the value of the observable under consideration at the physical pion mass. This is illustrated in the left panels of the plots in Figure \ref{fig:chiralCL}.

It remains to remove cut-off effects. Since our action is $\mathcal{O}(a)$-improved, the leading order cut-off effects present in our analysis are $\mathcal{O}(a^2)$, so the continuum limit is taken with the following ansatz:
\begin{equation}
  \mathcal{O}(a,m_\pi^\mathrm{phys},m_{D_s/\eta_c}^{\mathrm{ref},i},m_s^\mathrm{phys}) = \mathcal{O}(a=0,m_\pi^\mathrm{phys},m_{D_s/\eta_c}^{\mathrm{ref},i},m_s^\mathrm{phys}) + C_\mathcal{O}^\mathrm{CL}(m_{D_s/\eta_c}^{\mathrm{ref},i})\, a^2 + \mathcal{O}(a^4).
  \label{eq:CLansatz}
\end{equation}
As mentioned in the caption of Table \ref{tab:ensembles} the lattice spacing of the fine ensemble {\bf F1} is yet to be precisely determined. This is the origin of the large error bars on the green data points. The quantities in lattice units are very precisely determined, as can be seen from Figure \ref{fig:datacollection}.

\begin{figure}
  \centering
  \includegraphics[width=\textwidth]{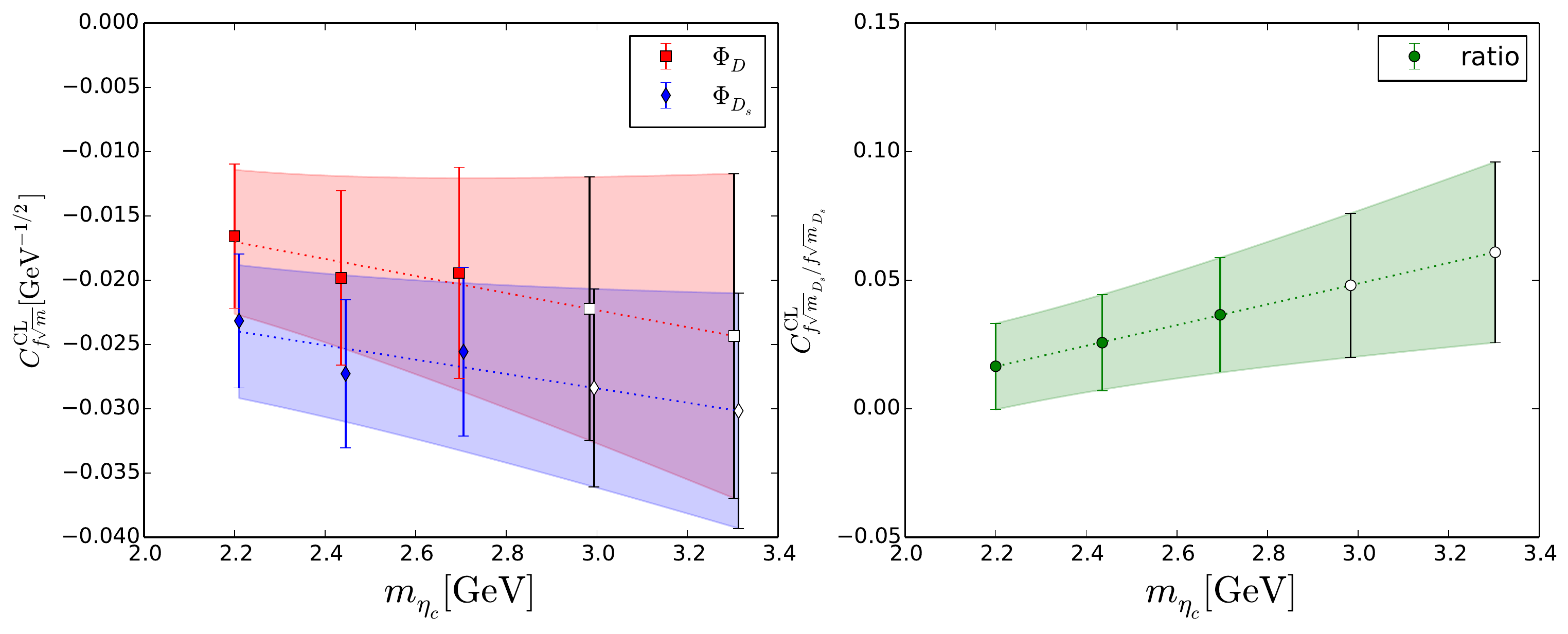}
  \caption{Parameterisation of the slope of the continuum limit for the example of $\Phi_D$ and $\Phi_{D_s}$ (left panel) as well as their ratio (right panel). The data points for $\Phi_{D_s}$ (blue diamonds) are slightly offset to the right for better visibility. The open black symbols correspond to the slopes of the continuum limit extrapolated from the region where this is well controlled.}
  \label{fig:CLparameterisation}
\end{figure}

The condition $am_h\leq 0.4$~\cite{Tsang:2014pfa} implies that on the coarsest ensemble we are unable to simulate the physical charm mass directly.
As a result of this, the continuum limit is not well constrained for reference masses $m_{D_s/\eta_c}^\mathrm{ref}$ above some maximal value $m_{D_s/\eta_c}^\mathrm{max}$, i.e. the heaviest reference mass for which we still have data on all three lattice spacings. 
 We also currently do not have a precise value for the lattice spacing on {\bf F1}, so the continuum limit is not well controlled for data where we only have 2 lattice spacings. 
However, the $C_\mathcal{O}^\mathrm{CL}$ should be a smooth function of $m_{D_s/\eta_c}^\mathrm{ref}$ for any given observable. We therefore parameterise the slope of the continuum limit $C_\mathcal{O}^\mathrm{CL}$ for values above $m_{D_s/\eta_c}^\mathrm{max}$ as
\begin{equation}
  C_\mathcal{O}^\mathrm{CL}(m_{D_s/\eta_c}^\mathrm{ref}) = C_\mathcal{O}^0 + C_\mathcal{O}^1 m_{D_s/\eta_c}^\mathrm{ref}.
  \label{eq:CLparameterisation}
\end{equation}
This parametrisation will also be helpful for a global fit ansatz, discussed in more detail in Section \ref{sec:summary}. The coefficients $C_\mathcal{O}^0$ and $C_\mathcal{O}^1$ are found from fitting the coefficients $C_\mathcal{O}^\mathrm{CL}(m_{D_s/\eta_c}^\mathrm{ref})$ in the region where the continuum limit is constrained by data on three lattice spacings. For $m_{D_s/\eta_c}^{\mathrm{ref},i} > m_{D_s/\eta_c}^\mathrm{max}$, the slope $C_\mathcal{O}^\mathrm{CL}(m_{D_s/\eta_c}^\mathrm{ref})$ is then constructed from (\ref{eq:CLparameterisation}) and the continuum limit fit (\ref{eq:CLansatz}) is carried out with this restriction. These slopes are shown for the example of $\Phi_D$, $\Phi_{D_s}$ (left) and their ratio (right) in Figure \ref{fig:CLparameterisation}. The black points are the continuum limit slopes obtained from this parameterisation.

\subsection{Extrapolation to Charm}
In the final step of the analysis we will interpolate the data to the physical value of the charm mass.
Several appraches are possible. One this to directly take the constrained continuum limit as described above for $m_h^{\mathrm{ref},i} = m_h^\mathrm{phys}$ (i.e. for $m_{\eta_c}^{\mathrm{ref},i} = m_{\eta_c}^\mathrm{phys}$ for the presented choice of interpolation variable). Alternatively one can recombine the results found for the different reference masses $m_h^{\mathrm{ref},i}$ using some fit form. The two different ways allow for an estimation of systematic errors. Finally, one could carry out the entire analysis choosing a different meson for the interpolation to fixed heavy quark masses, i.e. $m_{D_s}$ or $m_D$. This will also give some indication about systematic errors and the self-consistency of this approach. Since the analysis is still ongoing, we do not present extrapolated results here.

\section{Bag and $\xi$ Parameters Analyses}
Here we discuss the current fit strategies adopted in the analysis of the bag parameter and present preliminary results for the bare bag parameters and the ratio $\xi$.  
\subsection{Fit Strategies}
\begin{figure}
\centering
\includegraphics[width=0.7\textwidth]{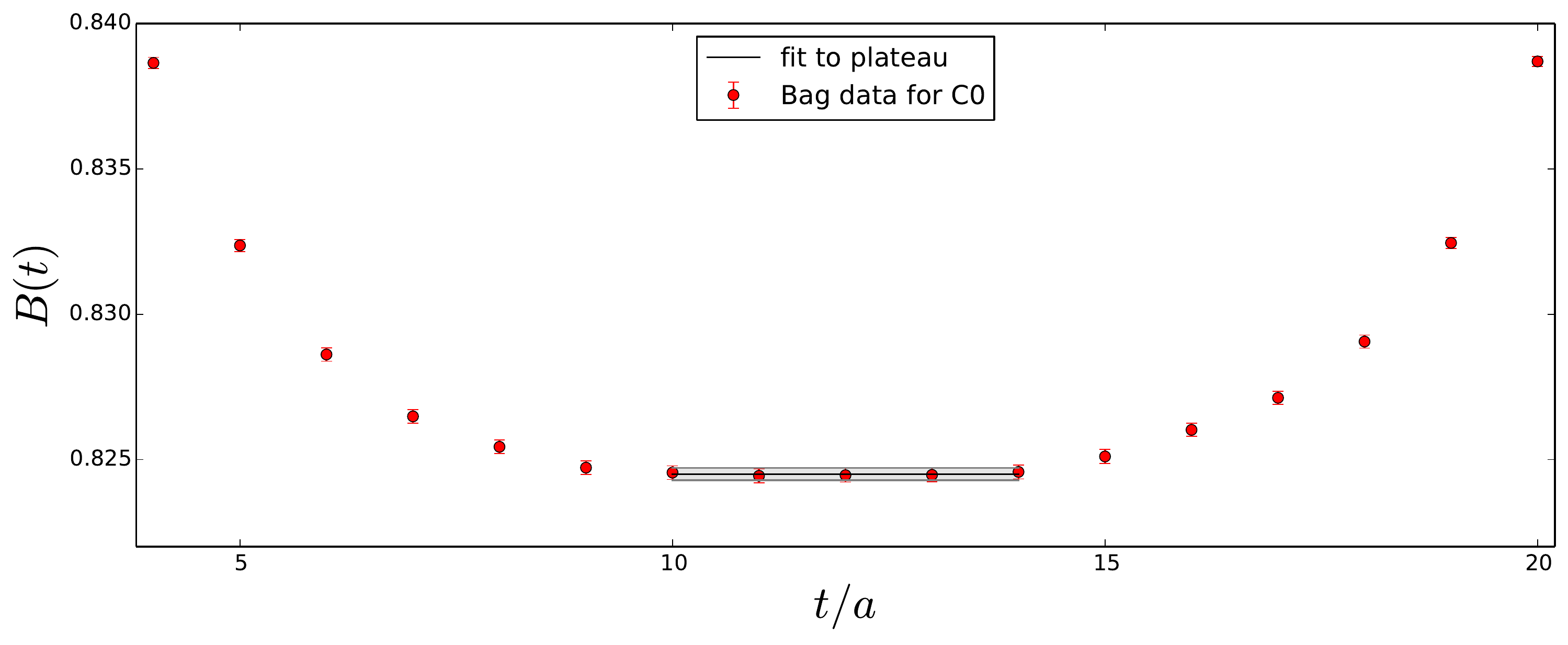}
\caption{Heavy-strange bag parameter on the ensemble {\bf C0} with $am_h=0.4$ as a function of time for $\Delta T/a=24$ and the fit to a constant when a plateau is reached.}
\label{bag-h0.4s-vs-t}
\end{figure}

\noindent For the bare bag parameter $B^\mathrm{bare}$ we fit the plateau to a constant in the region where the time dependence cancels 
\begin{equation}
B_P^\mathrm{bare}=\frac{\langle P^0(\Delta T)|O_{VV+AA}(t)|\bar{P}^0(0)\rangle}{\frac{8}{3}\langle P^0(\Delta T-t)|A_0(0)\rangle\langle A_0(t)|\bar{P}^0(0)\rangle}.
\end{equation}
This is performed for a number of different time separations between the walls $\Delta T/a$, where the mesons are located, and later combined in a simultaneous fit for statistical improvement. Generally small time separations suffer from excited states contaminations and a plateau is not reached, while for very large time separations the data becomes noisy. For example, on the {\bf C0} ensemble, we find $\Delta T/a$ in the range $20<\Delta T/a<30$ results in a plateau and small statistical error. Figure \ref{bag-h0.4s-vs-t} shows the heavy-strange bag parameter on the {\bf C0} ensemble. The heavy quark corresponds to $am_h=0.4$ as in Table~\ref{tab:ensembles}. The data in Figure~\ref{bag-h0.4s-vs-t} are obtained for time separation $\Delta T/a=24$.

\noindent Other strategies can be used to fit the data. For example, one can take into account the effect of the first excited state and hence extend the fit range. This is particularly useful for the case of heavy-light data since they become very noisy with increased $\Delta T/a$. \\

\noindent To obtain the $\xi$ parameter, the ratio of the results for the decay constants and the bag parameters for the heavy-strange and heavy-light mesons is computed using the jackknife scheme. The same analysis is then repeated for all four simulated charm masses on both lattices.

\subsection{Results}
\begin{figure}
\centering
\includegraphics[width=\textwidth]{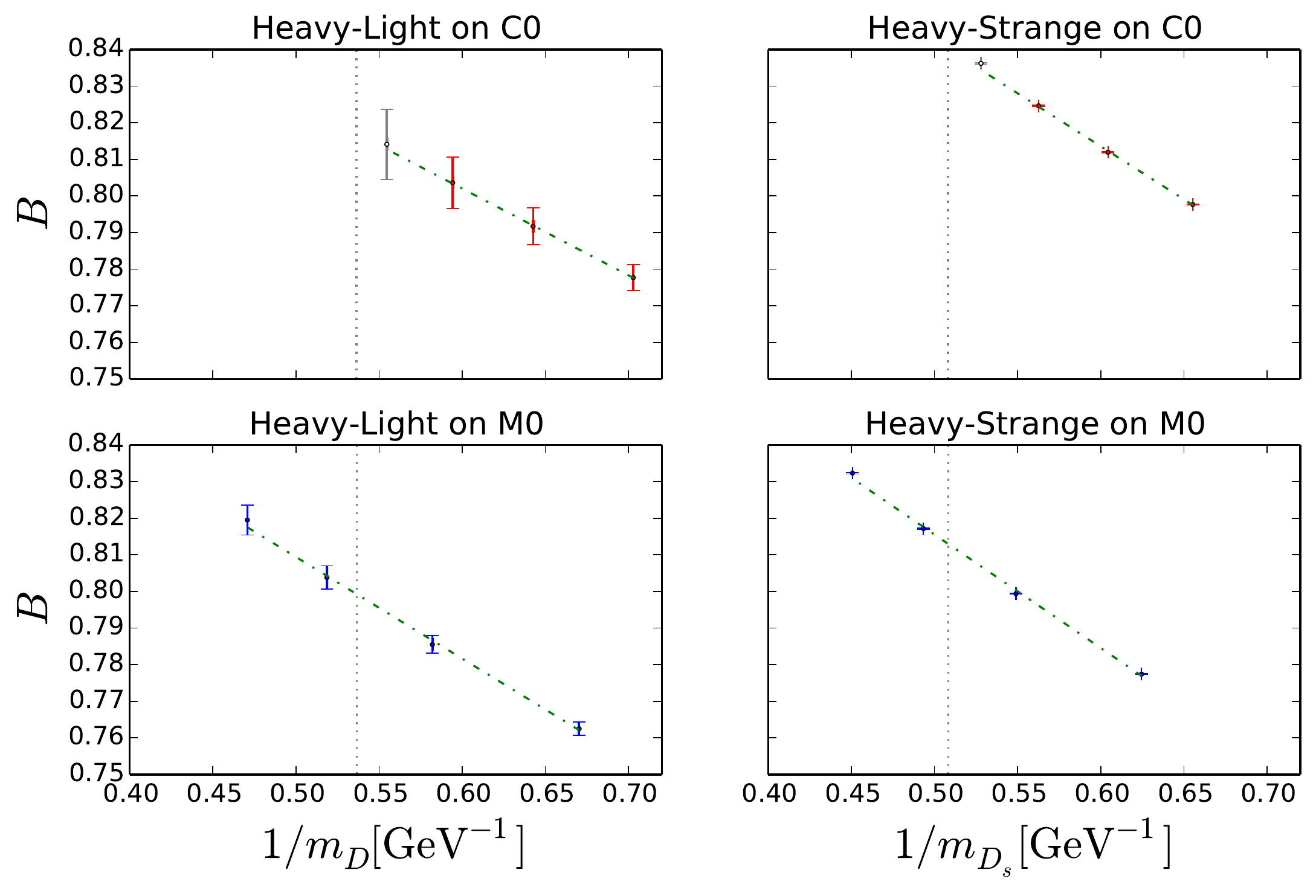}
\caption{Bag parameter $B$ vs. inverse meson mass for both heavy-light and heavy-strange mesons on both physical point lattices indicating linear behaviour. The dotted grey lines indicate inverse $D$ and  $D_s$ inverse meson masses. Note that the green dash-dotted line is simply drawn to guide the eye and this is not how we fit the bag parameter in practice. The heaviest point on the {\bf C0} ensemble, corresponding to $m_h=0.45$ is indicated in grey, as it does not enter the analysis. The details of this are explained in the text.}
\label{fig:bagvsinvmass}
\end{figure}

The results for heavy-light and heavy-strange bag parameters are plotted against inverse meson masses in Figure \ref{fig:bagvsinvmass}. The dotted grey lines correspond to $D$ and $D_s$ mesons. Presently, these are obtained using a single choice of time separation between the walls. As can be observed from the plots, the bag parameter depends linearly on inverse meson mass, suggesting that very few terms in an HQET expansion are required to describe our data at the current (percent scale) precision. Final conclusions are of course deferred until we have performed the mass and continuum extrapolations analyses. 
 
\noindent Note that the heaviest point on the {\bf C0} ensemble is shown by an open symbol because we suspect lattice artefacts to be significant.  Analysis of the midpoint correlation function $\langle J_{5q}(x)P(0)\rangle$, where $P$ is the pseudoscalar density and $J_{5q}$ is a pseudoscalar current mediating between the 5th dimension boundaries and the bulk, indicated poor binding of the DWF heavy quark fields to the walls. We will address this issue in more detail in Section.~\ref{sec:gaugesm}.\\

\begin{figure}
\centering
\includegraphics[width=0.8\textwidth]{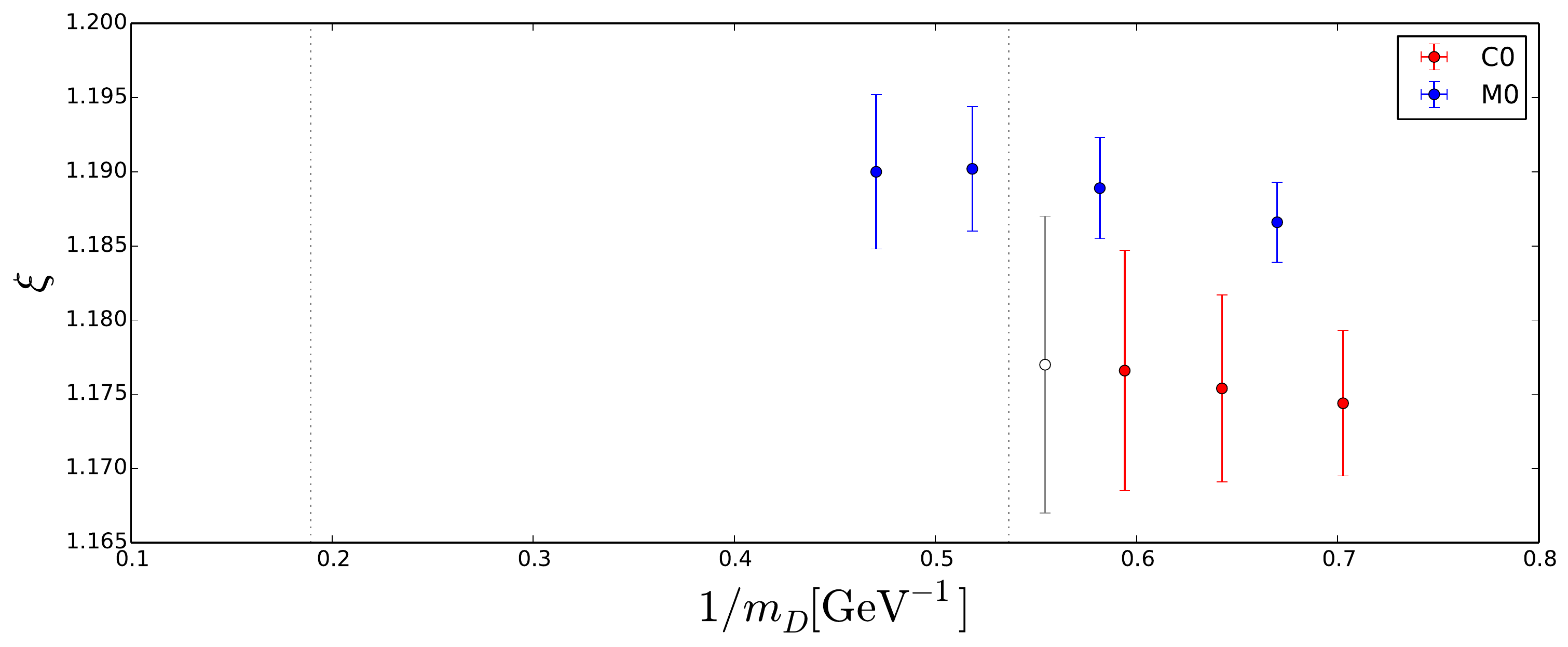}
\caption{$\xi$ parameter for heavy-light and heavy-strange mesons. The dotted grey lines indicate the inverse $D$ (right) and $B$ (left) meson masses.}
\label{fig:xiparamplot}
\end{figure}

The plot in Figure~\ref{fig:xiparamplot} shows the $\xi$-parameter vs inverse heavy-light meson mass. The dotted grey lines indicate inverse $D$ and $B$ meson masses to which an extrapolation will be made once the results from the third lattice spacing are available. 

Figure~\ref{fig:xiparamplot} shows remarkable insensitivity of $\xi$ to the heavy quark mass, perhaps even beyond the naive expectation from heavy quark effective theory since the data is consistent with very small $1/m_\mathrm{PS}$ terms and beyond at the present (small) statistical error. The largest theoretical uncertainty in $\xi$ to date has arisen from the chiral extrapolation \cite{Bazavov:2012zs,Gamiz:2009ku,Carrasco:2013zta}. We emphasise that these small errors have been obtained directly at physical pion masses which removes the need for such extrapolation. The dependence on the lattice spacing will be removed by a continuum extrapolation in a global fit in future work. Note that the {\bf M0} results have a higher precision compared to {\bf C0} in correspondence with that of the bag parameters. This is to be expected as an effect of greater self-averaging on the larger volume. 

\section{Gauge Link Smearing}\label{sec:gaugesm}
We have tested the effect of gauge link smearing on the axial current renormalization factor $Z_A$ with heavy-heavy quarks. The tests involve generating propagators on the {\bf A1} ensemble, with a different Stout smearing parameter $\rho$ and number of smearing hits \cite{Morningstar:2003gk}  as well as altering the domain wall height $M_5$ in the action \cite{Blossier:2009hg}. One can then check the effect of changing these parameters on the residual mass defined by
\begin{equation}
am_{res}=\frac{\sum_x\langle J_{5q}(x)P(0)\rangle}{\sum_x \langle P(x)P(0)\rangle},
\end{equation}
where $P$ is the pseudoscalar density and $J_{5q}$ is a pseudoscalar current mediating between the fifth dimension boundaries and the bulk.

\begin{figure}
\centering
\includegraphics[width=0.7\textwidth]{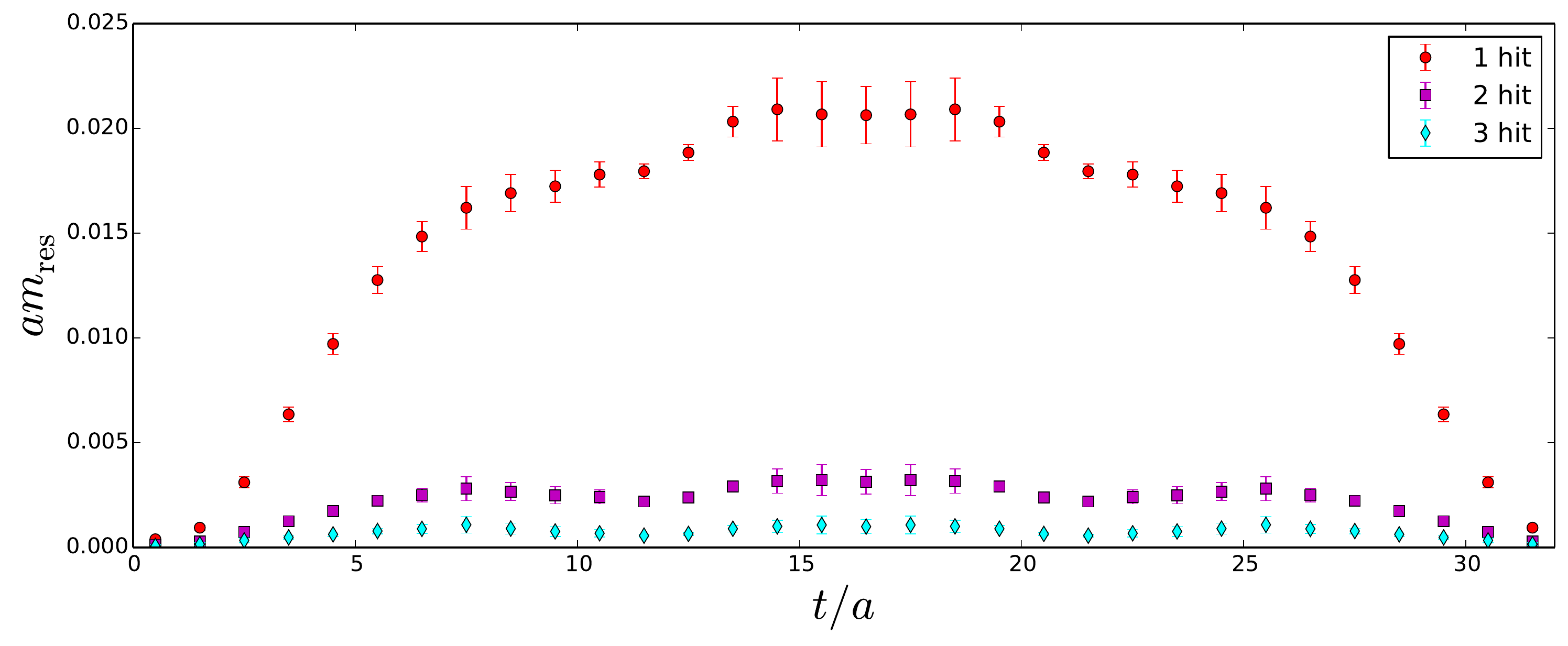}
\caption{Effect of different number of levels (hits) of Stout smearing on the residual mass with heavy quark input mass $am_h=0.45$ and $M_5=1.0$ carried out on the ensemble {\bf A1}.}
\label{mressmear}
\end{figure}

\noindent The purpose of the this study was to find the optimal heavy domain wall fermion action that would give access to heavier quark masses  i.e. closer to the physical point for $B$ physics studies. We seek minimum amount of smearing while still maintaining the light quark mass near its physical point. The simulated heavy quark mass is $am_h=0.45$ which is the one indicated by open symbols in the bag and $\xi$ parameter plots. Figure~\ref{mressmear} shows the effect of different number of Stout hits, with standard Stout parameter $\rho=0.1$. We observe that 3 hits of smearing reduces the residual mass to per mille level. Furthermore, it allows for simulation of even heavier masses whilst preserving the chiral properties of the domain wall formulation.

\section{Conclusion and Outlook}\label{sec:summary}
We have outlined our strategy for the predictions of our $D$ and $D_s$ phenomenology program using domain wall fermions for the light as well as the heavy propagators. We include two ensembles with near physical pion masses as well as three lattice spacings for a controlled $\mathcal{O}(a)$-improved continuum limit. The analysis strategy was presented using the decay constants as an example but the methodology will carry through to the other observables. The analysis is in an advanced state and we expect to publish predictions for first observables in the near future. The remaining work in progress mainly concerns the renormalization of the correlators and the precise and consistent determination of the lattice spacing on our fine ensemble. We need to estimate the systematic errors arising from choices in the presented analysis strategy.
We are also planning to carry out the whole analysis as one global fit to take all the small differences between the M\"obius and the Shamir ensembles into account. Finally, we hope to apply the ratio method~\cite{Blossier:2009hg} to make predictions for observables of the $B$ and $B_s$ mesons. 

At the same time, we are exploring changes in the formulation of the domain wall action, such as gauge link smearing, in order to increase the reach in the heavy quark mass. 
We will investigate the reach in heavy-light and heavy-strange meson masses using the parameters of the adapted action. This will potentially allow us to simulate directly at the charm quark mass on all three lattice spacings as well as reaching the heavier than charm region, allowing to better constrict the extrapolation to the $B$ sector.
The new action will then be used for the next large scale simulation of charm.




\acknowledgments{The research leading to these results has received funding from the European Research Council under the European Union's Seventh Framework Programme (FP7/2007-2013) / ERC Grant agreement 279757 as well as SUPA student prize scheme and STFC, grants ST/M006530/1, ST/L000458/1, ST/K005790/1, and ST/K005804/1, ST/L000458/1, and the Royal Society, Wolfson Research Merit Award, grantWM140078.
The authors gratefully acknowledge computing time granted through the STFC funded DiRAC facility (grants ST/K005790/1, ST/K005804/1, ST/K000411/1, ST/H008845/1).

\clearpage
{\small
\bibliographystyle{unsrt}    

\bibliography{library.bib}}



\end{document}